# Analysis of optical near-field energy transfer by stochastic model unifying architectural dependencies

Makoto Naruse,[1,2,a] Kouichi Akahane,[1] Naokatsu Yamamoto,[1] Petter Holmström,[3] Lars Thylén,[3,4] Serge Huant,[5] and Motoichi Ohtsu[2,6]

1 Photonic Network Research Institute, National Institute of Information and Communications Technology, 4-2-1 Nukui-kita, Koganei, Tokyo 184-8795, Japan

2 Nanophotonics Research Center, Graduate School of Engineering, The University of Tokyo, 2-11-16 Yayoi, Bunkyo-ku, Tokyo 113-8656, Japan

3 Laboratory of Photonics and Microwave Engineering, Royal Institute of Technology (KTH), SE-164 40 Kista, Sweden

4 Hewlett-Packard Laboratories, Palo Alto, CA 94304 USA

5 Institut Néel, CNRS and Université Joseph Fourier, 25 rue des Martyrs BP 166, 38042 Grenoble Cedex 9, France

6 Department of Electrical Engineering and Information Systems, Graduate School of Engineering, The University of Tokyo, 2-11-16 Yayoi, Bunkyo-ku, Tokyo 113-8656, Japan

a) Electronic mail: naruse@nict.go.jp

**Abstract:** We theoretically and experimentally demonstrate energy transfer mediated by optical near-field interactions in a multi-layer InAs quantum dot (QD) structure composed of a single layer of larger dots and $N$ layers of smaller ones. We construct a stochastic model in which optical near-field interactions that follow a Yukawa potential, QD size fluctuations, and temperature-dependent energy level broadening are unified, enabling us to examine device-architecture-dependent energy transfer efficiencies. The model results are consistent with the experiments. This study provides an insight into optical energy transfer involving inherent disorders in materials and paves the way to systematic design principles of nanophotonic devices that will allow optimized performance and the realization of designated functions.

# I. INTRODUCTION

Optical energy transfer via optical near-field interactions is one of the most important and unique attributes of nanophotonics.[1,2] Its theoretical fundamentals have been explained by local optical near-field interactions,[3,4] which describe optical energy transfer involving conventionally dipole-forbidden transitions, and the model predictions are in agreement with experimental demonstrations based on CdSe quantum dots (QDs),[5] ZnO quantum wells (QWs),[6] and ZnO QDs,[7] among others.[8,9] Higher-order multipolar interactions due to localized near-fields that break electric-dipole selection rules have also been discussed in the literature.[10-13] Optical energy transfer on the nanoscale has been applied to a variety of applications, including energy concentration,[14] logic circuits,[15] engineered color rendering for solid state lighting,[16] and computing paradigms beyond the conventional von Neumann architecture,[17] thanks to its unique attributes, such as the ability to break through the diffraction limit of light, high energy efficiency,[18] and spatiotemporal dynamics.[17]

What is important is to regulate the size and the position of nanostructures so that optical near-field interactions are induced between them in order to obtain designated functions.[19] This gives rise to the importance of modeling of nanophotonic devices and systems composed of multiple nanostructures arranged in varying configurations in characterizing and designing designated functions, such as efficient energy concentration,[20] wavelength conversion,[21] sensing,[22] color rendering for lighting and displays,[16] and many others. Technologies such as droplet epitaxy[23] and light-assisted quantum-dot size formation[24] have been developed, enabling geometry-regulated fabrication of nanostructured matter. At the same time, however, fabricated experimental devices exhibit more or less unavoidable inherent disorder,[25] such as variation of the sizes or layout of the nanostructures.

With regard to all of these requirements related to applications and issues originating from technological concerns, this paper provides a theoretical approach that takes into account the architecture and inherent disorder of nanostructures in a comprehensive manner. More concretely, we constructed a stochastic model for studying energy transfer via optical near-field interactions in multi-layer quantum dot devices, and we demonstrated that the results obtained with the model are consistent with experimental observations.

The concrete device that we studied is composed of multi-layer InAs QDs.[20] Akahane *et al.* have developed an original molecular beam epitaxy (MBE) technology for realizing multi-layer QD structures, where the size of the QDs in each layer and the inter-layer distances are precisely and individually controlled. Even QD devices with more than 300 layers have been realized with this technology.[26] Moreover, they have developed multi-layer QD devices in which one layer contains larger-sized QDs that are sandwiched by multiple layers of smaller-sized QDs.[20] More precisely, $N$ and $N+1$ layers (where $N$ is an integer) of smaller-sized QDs were located respectively below and above the larger-QD layer, as schematically shown in Fig. 1(a). Thanks to the optical energy transfer from smaller to larger QDs, the resulting light emission spectra differ depending on the number of smaller-sized QDs; that is to say, an $N$-dependence emerges. They experimentally demonstrated that the efficiency of energy transfer was maximized at a specific number of smaller QD layers. The presumable physical reasons behind this were mentioned in Ref. 20, but no precise modeling or quantitative discussions were given; this is an immediate typical case where a theoretical approach unifying device architectures, inherent disorder, and optical near-field interactions is necessary. This paper also expands on the experimental results reported in Ref. 20 by demonstrating layer-dependent photoluminescence.

This paper is organized as follows. Section II presents experimental results of energy transfer in stacked quantum dot devices and describes a basic theory of energy transfer via near-field interactions. Section III descibres stochastic modeling of stacked quantum-dot systems, the results of which are consistent with experiments. Section IV concludes the paper.

## II. ENERGY TRANSFER IN STACKED QUANTUM DOT

### A. Experiment

We begin by briefly reviewing some related studies. In the literature, the Forster model is a widely known model used for explaining optical energy transfer.[27] However, the issue of discrepancies between experimentally observed data and dipole-based models like the Forster model has been raised.[28,29] As mentioned above, optical near-field interactions take into account transitions that are conventionally dipole-forbidden,[3,10] and we have previously proposed a theory of a network of optical near-fields to describe a mixture of smaller and larger quantum dots.[19] Also, this theory has been successfully applied to a cascaded energy transfer device fabricated by a layer-by-layer chemical assembly method.[14,30] In these former studies, an idealized model yields good agreement with the experiments. We assume that the good agreement reported in the former studies is a consequence of the relatively simple device architectures used, and thus it is sufficient to characterize the dominant physical processes, which are inter-dot optical near-field interactions. In the case of much more complex devices, on the other hand, such as those composed of a large number of QDs, we need to take account of stochastic attributes inherent in the devices; for example, the model should satisfactorily explain fluctuations in quantum dot sizes, temperature-dependent energy level broadening, etc. At the same time, an exact theoretical treatment for large-scale systems, for example, with a density matrix formalism, results in an unnecessarily complex model, which will be computationally intractable in realistic computing environments. Therefore, keeping the model simple, while still extracting the essential physics, will be crucially important.[25]

Now we describe the experimental devices demonstrated in Ref. 21 and some newly fabricated devices, followed by their photoluminescence analysis. A multi-layer InAs QD structure was formed on InP(311)B based on MBE with a strain-compensation scheme.[31] As schematically shown in Fig. 1(a), relatively large-diameter QDs (average diameter: 47.9 nm, average height: 4.5 nm) were formed in the middle layer, and $N+1$ and $N$ layers of smaller-diameter QDs (average diameter: 42.8 nm, average height: 2.7 nm) were formed on the upper and lower sides, respectively. Six kinds of devices were fabricated, with $N$ equal to 1, 3, 5, 10, 15, and 20. The average inter-layer distance was 15 nm. Figure 1(b) shows normalized photoluminescence spectra, with vertical offsets, obtained by exciting the device with the 532 nm line of a YVO laser with a power of 120 mW at 300 K for the six devices. The spectra of devices with $N$ = 5, 10, and 20 have already been reported in Ref. 20, and the others were newly obtained here. In order to evaluate these six data sets on an equal basis, the photoluminescence spectra were calibrated so that the minimum signal levels, or noise floors, were aligned, and the spectra were normalized. Based on the spectra shown in Fig. 1(b), we evaluated the ratio of the photoluminescence from the larger QDs to that from the smaller QDs divided by the total number of such layers:

$$R = PL(L) / (PL(S)/\text{per layer}), \quad (1)$$

where PL(L) and PL(S) respectively denote photoluminescence from the larger and smaller QDs. We used three metrics to characterize PL(S) and PL(L). The solid curve in Fig. 1(c) shows the photoluminescence spectrum of the $N$ = 15 device, as an example. The first metric is the spectral peak values in shorter and longer wavelength ranges, which are respectively denoted by PL(S)$_p$ and PL(L)$_p$. The second one is the integrated photoluminescence intensity around the spectral peaks; here we integrate the intensities in the wavelength range ± 5 nm around the peaks, which are respectively denoted by PL(S)$_{p\pm 5\,nm}$ and PL(L)$_{p\pm 5\,nm}$. The last one is based on decomposing the spectrum into two Gaussian distributions; the dashed and dotted curves in Fig. 1(c) respectively indicate Gaussian profiles corresponding to larger and smaller QDs, and the calculated peaks of these profiles are taken as the metrics denoted by PL(L)$_f$ and PL(S)$_f$. The dash-dot curve in Fig. 1(c) represents the summation of the two Gaussian profiles, whose root-mean-square error to the original spectrum is minimized. We consider that this metric $R$ reflects the amount of optical energy transferred from the smaller dots to the larger ones. The circular, triangular, and square marks in Fig. 1(d) respectively show the metric $R$ based on the above three metrics evaluated with an excitation power of 120 mW at 300 K, as a function of $N$. The three metrics exhibited similar tendencies, where the maximum $R$ was obtained when $N$ was 15. In addition, Fig. 1(e) summarizes the metric $R$ as a function of the optical excitation power when $N$ was 15. The circular and square marks in Fig. 1(e) correspond to the cases where the temperature was 150 K and 300 K, respectively. As introduced at the beginning, in Ref. 21, Akahane *et al.* gave a qualitative discussion of the possible physics behind this behavior, including the effect of energy level broadening due to the temperature increase, but no quantitative reasoning or an explanation of the $N$-dependence was given. The goal of the present work is to reproduce such a tendency quantitatively and to identify one of the fundamental attributes of nanophotonics, as well as to provide a systematic stochastic method for use in designing and implementing optimized nanophotonic devices.

**B. Theoretical elements**

Now we focus on some theoretical elements of optical energy transfer via optical near-field interactions. The interaction Hamiltonian between an electron–hole pair and an electric field is given by

$$\hat{H}_{int} = -\int d^3r \sum_{i,j=e,h} \hat{\psi}_i^\dagger(\boldsymbol{r}) e\boldsymbol{r} \bullet \boldsymbol{E}(\boldsymbol{r}) \hat{\psi}_j(\boldsymbol{r}), \tag{2}$$

where $e$ represents the electron charge, $\hat{\psi}_i^\dagger(\boldsymbol{r})$ and $\hat{\psi}_j(\boldsymbol{r})$ are respectively electronic wave operators corresponding to the creation and annihilation operators of either an electron ($i, j = e$) or a hole ($i, j = h$) at $\boldsymbol{r}$, and $\boldsymbol{E}(\boldsymbol{r})$ is the electric field.[19] We consider a quantum dot based on a semiconductor material with a bulk electric dipole moment, on which a lightwave with a specific resonance frequency is incident. For the electronic system confined in a quantum dot, the electronic wave operator includes an envelope function of electronic waves in the quantum dot. In treating usual optical interactions of a quantum dot, the so-called long-wave approximation is employed, so that $\boldsymbol{E}(\boldsymbol{r})$ is considered to be constant over the size of the quantum dot since the electric field of propagating light wave is homogeneous on the nanometer scale. Depending on the symmetry of the electronic envelope function, numerical evaluation of the dipole transition matrix elements based on Eq. (2) results in selection rules for the optical transition for each excitonic state. For instance, in the case of spherical quantum dots, only optical transitions to the states specified by $l = m = 0$ are allowed, where $l$ and $m$ are the orbital angular momentum quantum number and magnetic quantum number for the envelope function, respectively. In the case of optical near-field interactions between a pair of resonant quantum dots separated by a sub-wavelength distance, on the other hand, due to the large spatial inhomogeneity of the optical near-fields of the source quantum dot, an optical transition that violates conventional optical selection rules becomes allowed, so that optical excitation transfer is possible between resonant states of quantum dots with different symmetries of the envelope functions via optical near-field interactions, regardless of the capability of each state to emit radiation into the optical far-field.[19]

Here we assume two spherical quantum dots whose radii are $R_S$ and $R_L$, which we call $QD_S$ and $QD_L$, respectively, as shown in Fig. 2 (a). The energy eigenvalues of states specified by quantum numbers $(n,l)$ are given by

$$E_{nl} = E_g + E_{ex} + \frac{\hbar^2 \alpha_{nl}^2}{2MR^2} \quad (n = 1, 2, 3, \cdots), \tag{3}$$

where $E_g$ is the band gap energy of the bulk semiconductor, $E_{ex}$ is the exciton binding energy in the bulk system, $M$ is the effective mass of the exciton, and $\alpha_{nl}$ are determined from the boundary conditions, for example, $\alpha_{n0} = n\pi, \alpha_{11} = 4.49$. According to Eq. (3), there exists a resonance between the level of quantum number (1,0) in $QD_S$ and that of quantum number (1,1) in $QD_L$ if $R_L / R_S = 4.49 / \pi \approx 1.43$. Note that optical excitation of the (1,1)-level in $QD_L$ corresponds to an electric dipole-forbidden transition. However, an optical near-field, denoted by $U$ in Fig. 2(a), allows this level to be populated due to the steep electric field in the vicinity of $QD_S$. Therefore, an exciton in the (1,0)-level in $QD_S$ could be transferred to the (1,1)-level in $QD_L$. In $QD_L$, the excitation undergoes intersublevel energy relaxation due to exciton–phonon coupling with a transition rate denoted by $\Gamma$, which is faster than the rate of the optical near-field interaction between the quantum dots,[30] and the excitation causes a transition to the (1,0)-level and radiation into the far-field. As a result, we find unidirectional optical excitation transfer from $QD_S$ to $QD_L$.

**III. STOCHSTIC MODELING**

The optical near-field interaction between two nanoparticles is known to be expressed as a screened potential using a Yukawa function, given by

$$U = \frac{A\exp(-\mu r)}{r}, \qquad (4)$$

where $r$ is the distance between the two dots, and the coefficient $\mu$ can take either real or imaginary values, which respectively correspond to localized and propagating modes.[3, 32] Here we assume a real-valued $\mu$ since we consider a localized mode. The detailed derivation of Eq. (4) is shown in Ref. 32.

In our stochastic modeling of the stacked QD devices, we take the following strategy. First, for the sake of simplicity while retaining the principal structure of the device, we assume that a larger dot, denoted by $L_0$, is located at the edge, and multiple smaller $QD_S$ are cascaded at one side of $L_0$, as schematically shown in Fig. 2(b). The smaller QDs are labeled $S_1, S_2, ..., S_N$, where $N$ is the total number of smaller QDs. Also, due to variation of the size of the QDs and energy level broadening due to temperature, the energy levels fluctuate and exhibit a certain width. The width of the energy band is denoted by $W_T$. The energy level deviation of a smaller QD with respect to the larger dot is denoted by $E_d$.

Here we consider that energy transfer from a smaller QD in the chain to the larger QD occurs when there exists an overlap between the resonant energy bands of the two QDs, as schematically shown in Fig. 2(a). We call such energy bands "energy-transfer-allowable" for the sake of later explanations. Note that we assume that the inter-dot energy transfer time is faster than the recombination time in individual dots. On the other hand, in the case where there is no energy band overlap, energy transfer does not take place; instead, the optical energy induced in the smaller dot contributes to the photoluminescence from the smaller dot, PL(S), as illustrated in Fig. 2(c). The optical energy transferred to the upper energy level of the larger dot relaxes to its lower energy band and results in photoluminescence from the larger dot, PL(L).

We assume that the sizes of the smaller QDs in the chain exhibit fluctuations that follow Gaussian statistics, which affects the energy band overlap. Take the example shown in Fig. 2(b), where the number of smaller dots is five ($N=5$), which are labeled $S_1, ..., S_5$. The dots $S_1$, $S_2$, and $S_5$ have energy-transfer-allowable energy bands; that is to say, they are resonant with the larger dot, $L_0$. Now, recall that the inter-dot energy transfer follows the Yukawa potential given by Eq. (4). We assume that energy transfer occurs perfectly if two adjacent dots have resonant energy bands; we represent such a situation by rewriting Eq. (4) such that a unit of energy induced in a smaller dot denoted by $i$ is transferred to the quantum dot denoted by $j$ with an efficiency given by

$$t_{i,j} = \frac{A'\exp(-\mu(d_{i,j}-d_C))}{d_{i,j}}, \tag{5}$$

where $d_C$ is the constant inter-layer distance between adjacent layers (which is about 15 nm, as mentioned above), and $d_{i,j}$ is the distance between QDs labeled $i$ and $j$. $A'$ is a constant so that the value $t_{i,j}$ becomes a dimensionless number, and is given by unity. This means, for example, that if two adjacent QDs have energy bands that are energy-transfer-allowable, $t_{i,j}$ is equal to 1 because $d_{i,j} = d_C$. Therefore, in the case shown in Fig. 2(b), the units of optical energy induced in S$_1$ and S$_2$ are both transferred to the larger dot, yielding radiation from the larger dot. Note that the optical energy induced in S$_2$ is first transferred to S$_1$ ($t_{S_2,S_1} = 1$), followed by a transfer from S$_1$ to L$_0$ ($t_{S_1,L_0} = 1$). The net amount of energy transferred to the larger dot from S$_2$ is $t_{S_2,S_1} \times t_{S_1,L_0} = 1$. The energy band of S$_5$ is also energy-transfer-allowable, meaning that a unit of optical energy induced in S$_5$ is transferred to L$_0$ by way of S$_2$ and S$_1$. Since the distance between S$_5$ and S$_2$ is $d_{S_5,S_2} = 3d_C$, the energy transferred from S$_5$ to S$_2$ is $t_{S_5,S_2} = \exp(-\mu(3d_C - d_C))/3d_C$, which is smaller than unity. The rest of the energy, $1 - t_{S_5,S_2}$, results in radiation from the smaller dot, which gives rise to PL(S). The energy $t_{S_5,S_2}$ transferred to S$_2$ is then transferred to S$_1$ with an efficiency $t_{S_2,S_1} = 1$. Summing up, the amount of radiation from the larger dot originating from S$_5$ is given by $\exp(-\mu(3d_C - d_C))/3d_C$. In this manner, we calculate the total amount of optical energy transferred to the larger QD and the radiation from the smaller QDs, which provides a metric corresponding to Eq. (1), that is, the amount of radiation from the larger QD divided by that from the smaller QDs, per layer.

We performed a Monte Carlo calculation based on the above modeling with 50,000 iterations for each of the devices with $N$ ranging from 1 to 20. We assumed that $\mu$ is unity and that the smaller QD size followed a Gaussian distribution with a standard deviation of unity, and we equated such a distribution with the energy level deviation from the larger QD. The energy band of the larger QD is represented by width $W_T$, with its center being equal to the mean of the normal distribution. The energy band of a smaller QD corresponds to a range whose width is also $W_T$, and its center is displaced from the mean of the normal distribution by $E_d$, as schematically shown in Fig. 3(a).

The square, circular, and triangular marks shown in Fig. 3(b) represent the ratio $R$ as a function of the number of smaller QD layers ($N$) when $W_T$ was 0.5, 1, and 1.5, respectively. The profile corresponding to the case where $W_T$ was 1.5 exhibits a similar tendency to the experimental results shown in Fig. 1(d), which were obtained at 300 K. In the experiment, there exists a maximum value $R$ of around 16 when $N$ is 15, whereas in the calculation, the maximum $R$ is 12.1 when $N$ is 10. We consider that these results show consistency between the experiment and the stochastic model.

To further evaluate the validity of the model, assume that the inter-dot interaction does *not* follow the Yukawa potential; specifically, suppose that the energy transfer efficiency does not depend on the inter-dot distance. All of the other assumptions are the same as the previous ones. In such a case, the energy transfer performance depending on the number of layers results in the profiles shown in Fig. 3(c), where the ratio $R$ monotonically increases as a function of $N$, which is not consistent with the experiment. This is another indication that the inter-layer-distance–dependent near-field interactions, as well as the inherent stochastic distributions, are the origin of the optimized energy transfer observed in the multi-layer QD device.

The temperature- and excitation-power–dependent energy transfer experimentally demonstrated in Fig. 1(e) are not completely manageable in the proposed stochastic model. The temperature-dependence, however, is partially reproduced by the model. In the modeling results demonstrated in Fig. 3(b), the ratio $R$ exhibits a larger value when $W_T$, that is, the temperature, increases. This is consistent with the experimental results shown in Fig. 1(e), where the energy transfer efficiencies at the higher temperature (300 K) are larger than those at the lower temperature (150 K). There was an exception in Fig. 1(e) when the excitation power was high (100 mW), where the low-temperature experiment yielded a larger value than the high-temperature one. More detailed modeling will be needed to fully explain this, for example, by including state filling effects which may occur in cases where the excitation power is high.[19] This will be one of the topics of future work.

## IV. SUMMARY

In summary, we demonstrated energy transfer mediated by optical near-field interactions in a multi-layer InAs quantum dot (QD) structure composed of a single layer of smaller dots and *N* layers of larger ones and examined its basic principles by using a stochastic model in which optical near-field interactions that follow a Yukawa function, QD size fluctuations, and temperature-dependent energy level broadening are unified. The properties of experimentally observed energy transfer and the results calculated from the model were in good agreement. This study provides an insight into optical energy transfer while taking account of inherent disorder and paves the way to systematic design principles of nanophotonic devices for achieving optimized performance and realizing designated functionalities.

## ACKNOWLEDGEMENTS

This work was supported in part by the Japan–Sweden Bilateral Joint Project supported by the Japan Society for the Promotion of Science (JSPS). This work was also supported in part by the Strategic Information and Communications R&D Promotion Programme (SCOPE) of the Ministry of Internal Affairs and Communications, and by Grants-in-Aid for Scientific Research from the Japan Society for the Promotion of Science.

**References**

1. M. Ohtsu, K. Kobayashi, T. Kawazoe, S. Sangu, and T. Yatsui, IEEE J. Sel. Top. Quantum. Electron. **8**, 839 (2002).

2. V. I. Klimov ed., *Semiconductor and Metal Nanocrystals* (Marcel Dekker, New York, 2003).

3. K. Kobayashi, S. Sangu, H. Ito and M. Ohtsu, Phys. Rev. A **63**, 013806 (2000).

4. R. Filter, S. Mühlig, T. Eichelkraut, C. Rockstuhl, and Falk Lederer, Phys. Rev. B **86**, 035404 (2012).

5. R. C. Somers, P. T. Snee, M. G. Bawendi, , D. G. Nocera, J. Photochemistry and Photobiology A: Chemistry **248**, 24 (2012).

6. T. Yatsui, S. Sangu, T. Kawazoe, M. Ohtsu, S. J. An, J. Yoo, and G.-C. Yi, Appl. Phys. Lett. **90**, 223110 (2007).

7. T. Yatsui, H. Jeong, and M. Ohtsu, Appl. Phys. B **93**, 199 (2008).

8. P. Vasa, R. Pomraenke, S. Schwieger, Yu. I. Mazur, Vas. Kunets, P. Srinivasan, E. Johnson, J. E. Kihm, D. S. Kim, E. Runge, G. Salamo and C. Lienau, Phys. Rev. Lett. **101**, 116801 (2008).

9. R. C. Somers, P. T. Snee, M. G. Bawendi, Daniel G. Nocera, J. Photochem. Photobio. A: Chem. **248**, 24 (2012).

10. J. R. Zurita-Sánchez and L. Novotny, J. Opt. Soc. Am. B **19**, 1355- 1362 (2002).

11. J. R. Zurita-Sánchez and L. Novotny, J. Opt. Soc. Am. B **19**, 2722- 2726 (2002).

12. O. Mauritz, G. Goldoni, F. Rossi, and E. Molinari, Phys. Rev. Lett. **82**, 847 (1999).

13. R. Filter, S. Mühlig, T. Eichelkraut, C. Rockstuhl, and F. Lederer, Phys. Rev. B **86**, 035404 (2012).

14. T. Franzl, T. A. Klar, S. Schietinger, A. L. Rogach, and J. Feldmann, Nano Lett. **4**, 1599‑1603 (2004).

15. C. Pistol, C. Dwyer, and A. R. Lebeck, IEEE Micro **28**, 7-18 (2008).

16. B. A. Akins, G. Medina, T. A. Memon, A. C. Rivera, G. A. Smolyakov, and M. Osinski, "Nanophosphors Based on CdSe/ZnS and CdSe/$SiO_2$ Colloidal Quantum Dots for Daylight-Quality White LEDs," in *Conference on Lasers and Electro-Optics 2010*, OSA Technical Digest (CD) (Optical Society of America, 2010), paper CTuNN7.

17. M. Aono, M. Naruse, S.-J. Kim, M. Wakabayashi, H. Hori, M. Ohtsu, M. Hara, Langmuir **29**, 7557-7564 (2013).

18. M. Naruse, H. Hori, K. Kobayashi, P. Holmström, L. Thylén, and M. Ohtsu, Opt. Express **18**, A544 (2010).

19. M. Naruse, N. Tate, M. Aono, and M. Ohtsu, Rep. Prog. Phys. **76**, 056401 (2013).

20. K. Akahane, N. Yamamoto, M. Naruse, T. Kawazoe, T. Yatsui, and M. Ohtsu, Jpn. J. Appl. Phys. **50**, 04DH05 (2011).

21. I. Robel, V. Subramanian, M. Kuno, and, P. V. Kamat, J. Am. Chem. Soc. **128**, 2385 (2006).

22. V. Bagalkot, L. Zhang, E. Levy-Nissenbaum, S. Jon, P. W. Kantoff, R. Langer, and O. C. Farokhzad, Nano Letters **7**, 3065-3070 (2007).

23. T. Mano, T. Kuroda, K. Mitsuishi, Y. Nakayama, T. Noda and K. Sakoda, Appl. Phys. Lett. **93**, 203110 (2008).

24. H. Koyama and N. Koshida, J. Appl. Phys. 74, 6365 (1993).

25. S. E. Skipetrov, Phys. Rev. E **67**, 036621 (2003).

26. K. Akahane, N. Yamamoto, and T. Kawanishi, physica status solidi (a) **208**, 425 (2011).

27. T. Förster, Ann. Phys. **2**, 55 (1948).

28. G. D. Scholes and G. R. Fleming, J. Phys. Chem. B **104**, 1854 (2000).

29. M. Kubo, Y. Mori, M. Otani, M. Murakami, Y. Ishibashi, M. Yasuda, K. Hosomizu H. Miyasaka, H. Imahori, and S. Nakashima, J. Phys. Chem. A **111**, 5136 (2007).

30. M. Naruse, E. Runge, K. Kobayashi, and M. Ohtsu, Phys. Rev. B **82**, 125417 (2010).

31. K. Akahane, N. Ohtani, Y. Okada, and M. Kawabe, J. Cryst. Growth **245**, 31 (2002).

32. S. Sangu, K. Kobayashi, A. Shojiguchi, T. Kawazoe, and M. Ohtsu, J. Appl. Phys. **93**, 2937 (2003).

**Figure captions**

**FIG. 1.** (Color online) (a) A stacked quantum dot (QD) structure in which a large-sized QD layer is sandwiched by $N$ and $N+1$ layers of smaller-sized QDs. (b) $N$-dependent photoluminescence spectra which differ due to energy transfer from smaller to larger QDs. (c) Photoluminescence spectrum of the device $N=15$ (solid curve) and two Gaussian profiles assumed for radiation from smaller QDs (dotted curve) and larger QDs (dashed curve). The dash-dot curve represents the summation of the two Gaussian profiles. (d) The metric $R$, which is the photoluminescence intensity from the larger-dot layer divided by that from the smaller-dot layers, per layer. The circular-, triangular-, and square-marks are based on different metrics defined in (c). See main text for detailed definitions. (e) The metric $R$ evaluated at different temperatures and different excitation powers.

**FIG. 2.** (Color online) (a) Schematic diagram of optical energy transfer from a smaller QD to a larger one via an optical near-field interaction ($U$). The energy levels are broadened to width $W_T$ due to temperature. When there is an overlap between the energy bands, energy transfer is induced. (b) Schematic diagram of a larger dot ($L_0$) and an array of $N$ cascaded smaller dots ($S_1$, $S_2$, ..., $S_N$). The example shows the case where $N=5$. The optical energy induced in "energy transfer allowable" bands, which are in $S_1$, $S_2$, and $S_5$, is transferred to $L_0$ by way of such dots and contributes to the photoluminescence from the larger dot (PL(L)). Photoluminescence from the smaller dots is represented by PL(S). (c) Optical energy induced in the smaller dots that are not resonant with the larger one is not delivered to the larger dot.

**FIG. 3.** (Color online) (a) The size of the smaller QDs, that is, the relative energy level difference with respect to the larger QD, is assumed to follow a normal distribution. The energy level broadening and the relative difference with respect to the larger dot are respectively denoted by $W_T$ and $E_d$. (b) The metric $R$ as a function of the number of smaller QDs provided in the stochastic model. In the case where $W_T$ is 1.5, the profile and the maximum $R$ give results consistent with the experimental demonstration summarized in Fig. 1(d). Also, we observed that as $W_T$ increases, the metric $R$ increases; this is consistent with the temperature dependence experimentally observed in Fig. 1(e). (c) When the inter-dot optical near-field is assumed not to follow the Yukawa potential, $R$ simply linearly increases as $N$ increases, which is not consistent with the experiment.

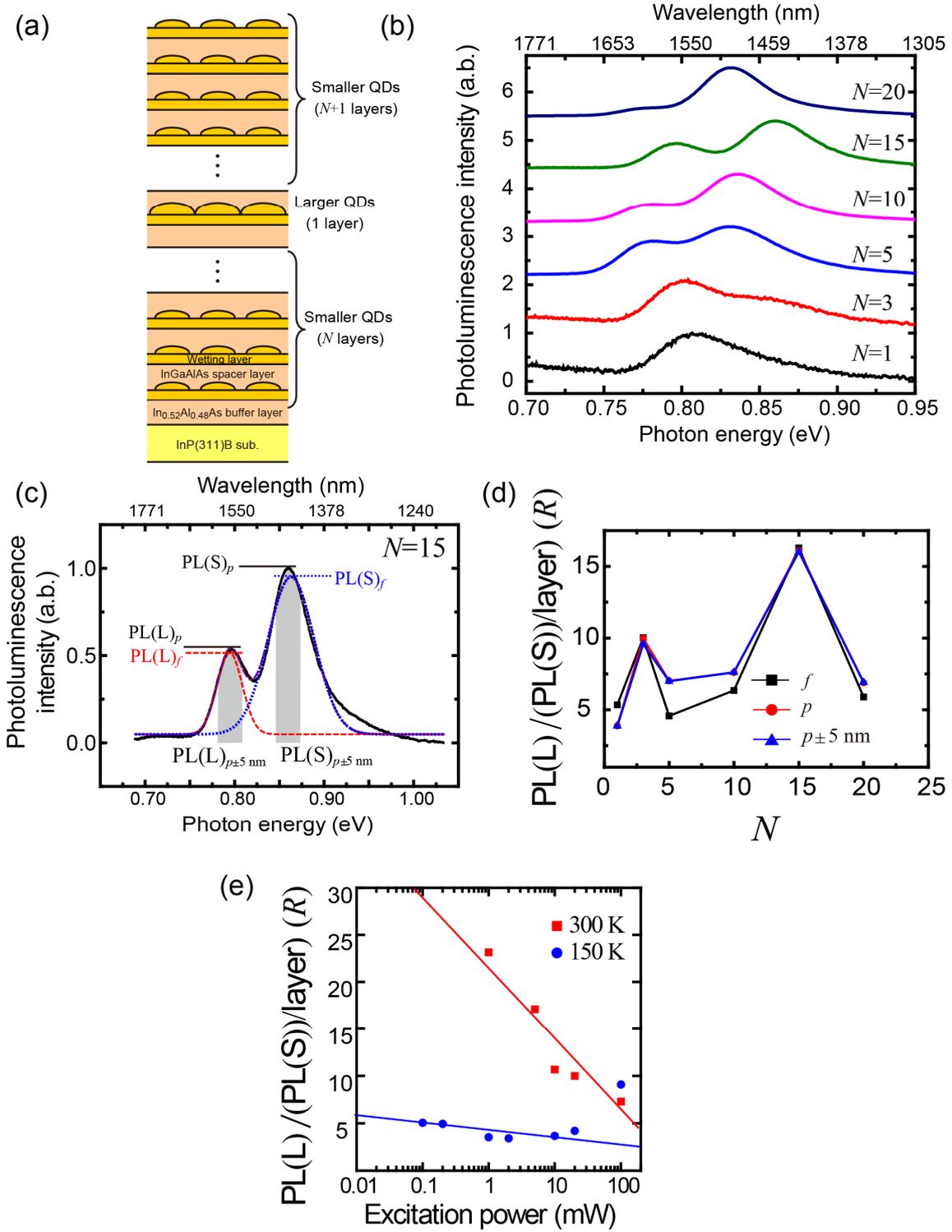

FIG. 1

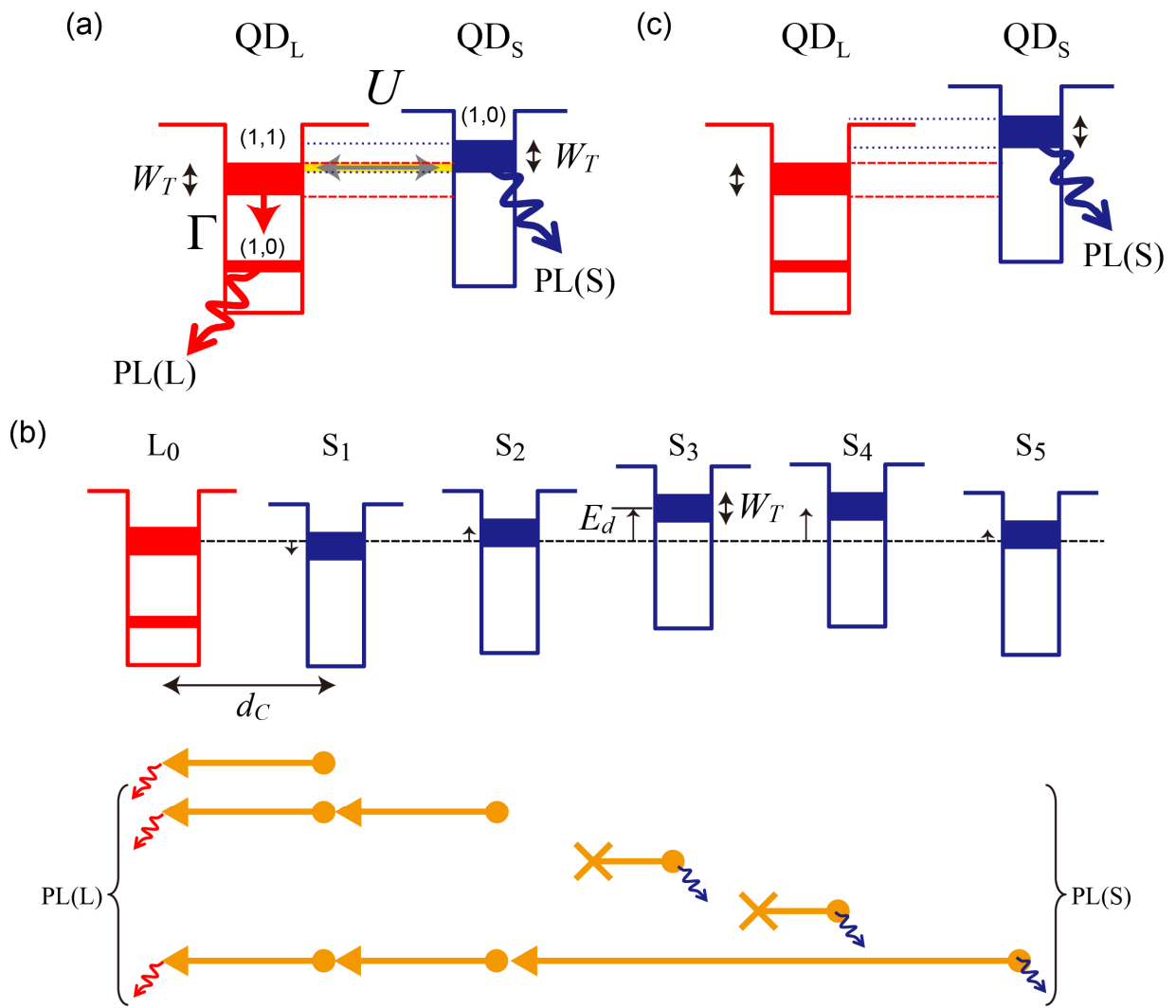

FIG. 2

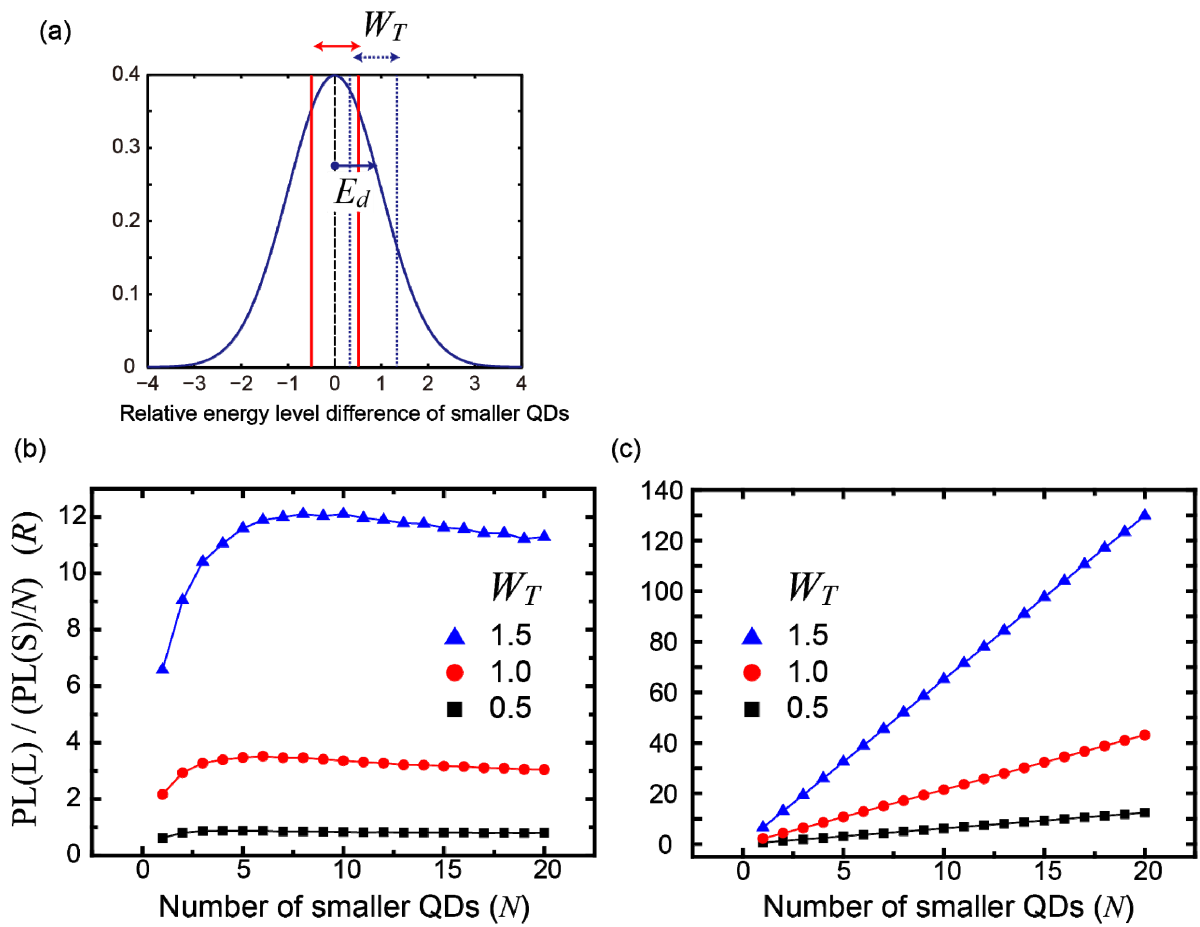

FIG. 3